# Automated monitoring of bee colony movement in the hive during winter season

Rostyslav Koroliuk[1,†], Vyacheslav Nykytyuk[1,†], Vitaliy Tymoshchuk[1, *,†], Veronika Soyka[2,†] and Dmytro Tymoshchuk[1,†]

[1] *Ternopil Ivan Puluj National Technical University, Ruska str. 56, Ternopil, 46001, Ukraine*

[2] *Ternopil Volodymyr Hnatiuk National Pedagogical University, Maxyma Kryvonosa str. 2, Ternopil, 46027, Ukraine*

**Abstract**
In this study, we have experimentally modelled the movement of a bee colony in a hive during the winter season and developed a monitoring system that allows tracking the movement of the bee colony and honey consumption. The monitoring system consists of four load cells connected to the RP2040 controller based on the Raspberry Pi Pico board, from which data is transmitted via the MQTT protocol to the Raspberry Pi 5 microcomputer via a Wi-Fi network. The processed data from the Raspberry Pi 5 is recorded in a MySQL database. The algorithm for finding the location of the bee colony in the hive works correctly, the trajectory of movement based on the data from the sensors repeats the physical movement in the experiment, which is an imitation of the movement of the bee colony in real conditions. The proposed monitoring system provides continuous observation of the bee colony without adversely affecting its natural activities and can be integrated with various wireless data networks. This is a promising tool for improving the efficiency of beekeeping and maintaining the health of bee colonies.

## 1. Introduction

The popularity of honeydew honey is growing among consumers and producers alike due to its medicinal properties, which are better than those of most flower honey [1]. Studies show that some types of honeydew honey help to suppress antibiotic-resistant bacteria [2]. However, it has a negative effect on the bees themselves, as the composition of honeydew honey is rich in trisaccharide mellitosis, which is poorly digested and can cause malnutrition and death in bees [3]. Therefore, it is important for beekeepers to have comprehensive information about the bee colony: its condition, position in the hive, food supply, humidity, temperature, etc. Electromechanical devices or handwritten records have often been used to obtain bee colony data, for example, the authors in [4] studied bee behavior by marking bees and manually counting them as they appeared at feeders. Although such studies have provided a lot of valuable data, their ability to scale to multiple hives and the ability to monitor bees in the field is severely limited. These methods require a large amount of work, both at the initial stage and during the acquisition of the necessary information.

With the spread of information technology, the possibilities for researching bee colonies have also expanded. Using temperature [5-8], audio [9], video or radiation [10-14], and artificial intelligence [15], it was possible to obtain indicators of bee vital activity without performing a large amount of mechanical work.

Given the successful implementation of monitoring technologies in hives, it became clear that the development of 'smart hives' combining two or more sensors was inevitable. In particular, the authors of [16] used a wireless sensor network (WSN) to measure hive parameters: $CO_2$, $O_2$, pollutant gases (nitrogen dioxide ($NO_2$), ethanol ($CH_3CH_2OH$), ammonia ($NH_3$), carbon monoxide (CO) and methane ($CH_4$)), temperature and relative humidity. The analysis of these data provided a unique picture of the state of bee hives in unfavorable conditions (night, bad weather). In addition, the authors in [17, 18] installed a camera for photos and videos, a USB microphone, a humidity and temperature sensor in their Beemon system. In [19], a hive control system is built on the basis of WSNs and microservices to monitor parameters inside the hive and the environment at the hive location. In [20], the monitoring of temperature, humidity (inside and outside the hive), hive weight, and sound inside the hive was chosen, as these are the most frequently mentioned indicators for assessing the health of bee colonies. An additional mechanical sensor was also included to detect the opening of the hive lid, for example, as a result of an animal attack or strong winds.

The aim of our work is to experimentally model the movement of a bee colony in a hive during the winter season and develop an inexpensive and efficient monitoring system that allows tracking the movement of the bee colony and honey consumption without interfering with the natural activities of the bee colony.

## 2. Materials and methods

To track the position of bees in the hive during the winter season, we decided to use the load cells, as photo and video equipment can interfere with the work of bees and is quite expensive. We designed and manufactured a hive stand with adjustable legs and a platform under the hive bottom with four load cells in the corners (Figure 1).

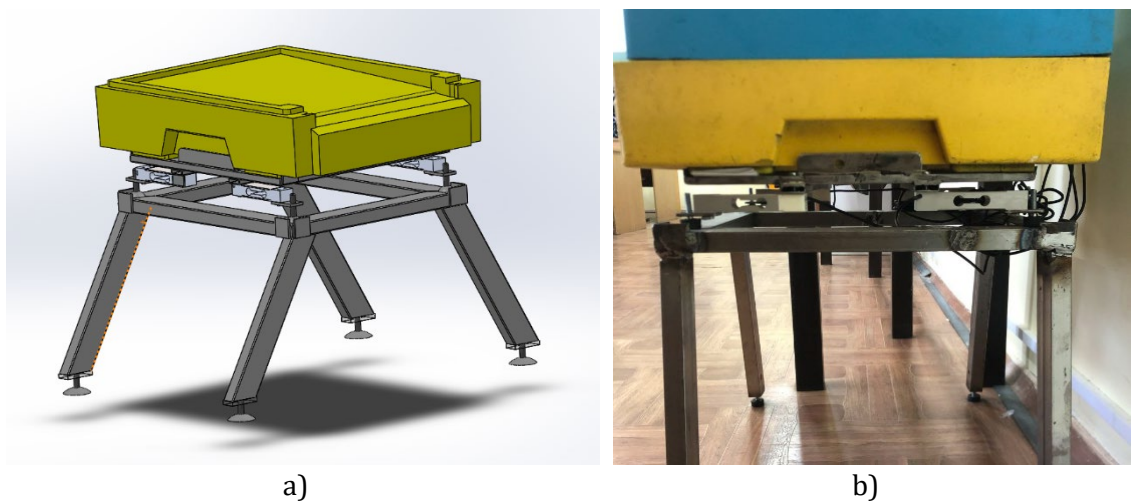

a)  b)

**Figure 1:** Hive stand with measuring platform and hive bottom a) 3D model; b) finished product.

The load cells allow measuring and recording any changes in weight distribution within the hive structure, which makes it possible to obtain data for developing a model of bee family movement. To reduce "noise", the readings were made ten times, and the average value of each load cell was calculated using formula (1) and stored in the experimental database.

$$param = \frac{\sum_{i=1}^{10} data_i}{10} \qquad (1)$$

where *param* is the average value of the load cell indicators, $data_i$ is the current load cell indicator measurements.

When each load cell is placed at the corners of the measuring platform, the sum of the forces from all 4 load cells is equal to the force applied by the weight of the platform and objects on it [21, 22]. In order to find the point at which an object is located or a force is applied, it is necessary to enter a coordinate system.

Suppose that a certain force *F* is applied at point *H* (x, y) on a platform of size *n* x *m* and that each load cell is subject to forces *F1*, *F2*, *F3*, *F4*. Define a coordinate system with the origin at load cell #1 (Figure 2).

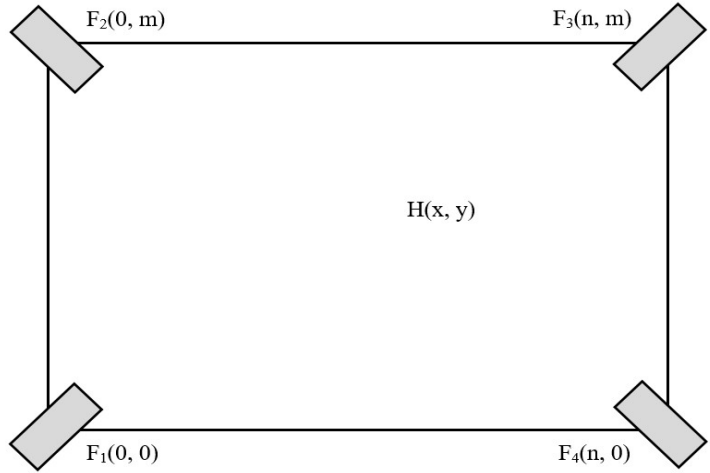

**Figure 2:** Coordinate system of load cells.

Let us use the formula for finding the center of mass of a system of material points [23]:

$$\vec{r_c} = \frac{\sum_{i=1}^{k} m_i \vec{r_i}}{\sum_{i=1}^{k} m_i} \qquad (2)$$

where $\vec{r_c}$ is the radius vector of the centre of mass, $m_i$ is the mass of the *i*-th material point, $\vec{r_i}$ is the radius vector of the *i*-th material point, *n* is the number of material points in the system.

The center of mass is a weighted average of the radius vectors of the points, where the weights are the masses of these points.

So we write (2) with our data:

$$\vec{r_c} = \frac{\sum_{i=1}^{4} F_i \vec{r_i}}{\sum_{i=1}^{4} F_i} = \frac{F_1 \vec{r_1} + F_2 \vec{r_2} + F_3 \vec{r_3} + F_4 \vec{r_4}}{F_1 + F_2 + F_3 + F_4} = \frac{F_1 \vec{r_1} + F_2 \vec{r_2} + F_3 \vec{r_3} + F_4 \vec{r_4}}{F} \qquad (3)$$

where $\vec{r_c}$ is the radius vector of the center of mass of the system of material points, $F_i$ is the weight of the object (force measured by the load cell) of the *i*-th material point, $\vec{r_i}$ is the radius vector of the *i*-th material point, i.e. the position of the point on the plane, $F$ is the sum of the forces in the system.

The expression is simplified by using specific coordinates for each point:

$$\vec{r_c} = \frac{F_1(0 \cdot \vec{i} + 0 \cdot \vec{j}) + F_2(0 \cdot \vec{i} + m \cdot \vec{j}) + F_3(n \cdot \vec{i} + m \cdot \vec{j}) + F_4(n \cdot \vec{i} + 0 \cdot \vec{j})}{F}$$
$$= \frac{(F_3 + F_4)n}{F}\vec{i} + \frac{(F_2 + F_3)m}{F}\vec{j} \qquad (4)$$

where $\vec{r_c}$ is the radius vector of the centre of mass of the system of material points, *i* is a unit vector along the *x*, *j* is a unit vector along the *y*, *n* is the distance indicating the position of the point along the *x*-axis, *m* is the distance indicating the position of the point along the *y*, *F* is the sum of the forces in the system.

From equation (3), we obtain equations (4) and (5) to determine the coordinates *x* and *y* of the center of mass (center of gravity) of the system of material points:

$$x = \frac{(F_3 + F_4)n}{F} \quad (5)$$

$$y = \frac{(F_2 + F_3)m}{F} \quad (6)$$

where *x* is the coordinate of the center of mass along the *x*-axis, *y* is the coordinate of the center of mass along the *y*-axis.

Knowing the forces and the total size of the measuring platform, it is possible to calculate the center of mass of the bee colony using equations (5) and (6).

Experimental modelling of bee colony movement in a hive in the winter season was carried out according to the data on bee colony movement and average honey consumption in winter described in [24-26].

The trajectory of bee colony movement mainly goes from the entrance to the hive upwards and then horizontally to the back wall of the hive (Figure 3).

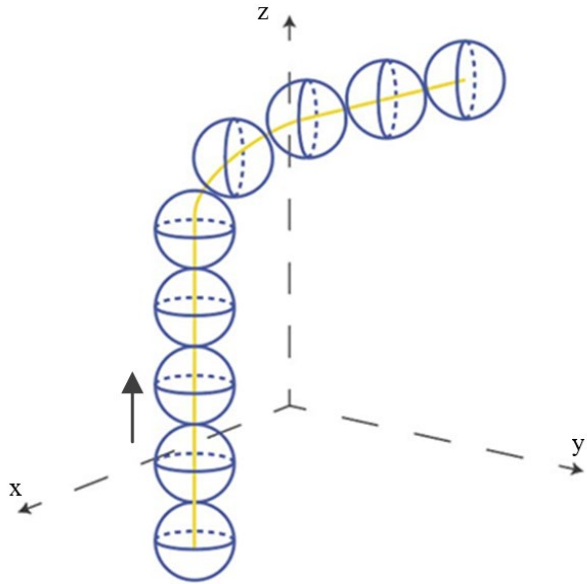

**Figure 3:** The trajectory of a bee colony in a hive during the winter season.

To simulate the consumption of honey by a bee colony in winter, a model of filling a 10-frame hive with honey before wintering was used. Metal strips with holes were used to simulate honey filling. The strips were 4 mm thick and 355 mm, 185 mm, and 35 mm long. The total weight of the strips is 30 kg. The weight of the smallest strips (26-27 g) roughly corresponds to the daily honey consumption by the swarm in early winter.

The bee's swarm was simulated with a sphere 200 mm in diameter and weighing 2 kg, which was divided into 6 parts and printed on a 3D printer. This corresponds roughly to a bee family of 20,000 individuals.

# 3. Results

The following steps were taken to reproduce the actual behavior of the bee colony and to read the data. First, the hive stand was set up and levelled using the adjustable feet. After that, the measuring platform was placed. Next, we connected the load cells to the RP2040 controller and performed the taring. A beehive with a simulated honey filling was placed on the platform, and a sphere simulating a bee swarm was fixed with a special frame.

Next, the sphere was moved 1 mm upwards and the smallest plate under the sphere was removed, after which the load cells were read and the average values were recorded in the experiment database. This process was repeated 70 times, simulating the movement of a bee colony during the first 70 days of wintering. Then the sphere was moved 2 mm horizontally and the three smallest plates under the sphere were removed, the load cells were read again and the average values were recorded in the experiment database, which corresponded to the movement and consumption of honey by the swarm over two days. This procedure was repeated 25 times, modelling the next 50 days of bee colony movement.

In the last step, the sphere was moved horizontally by 1 mm and the two smallest plates under the sphere were removed, after which the load cells were again measured and the average values were recorded in the database. This process corresponded to the movement and consumption of honey by the swarm in one day in the final phase of wintering. This step was repeated 50 times to complete the experiment.

Figure 4 shows a laboratory diagram of the bee colony movement monitoring system in the hive.

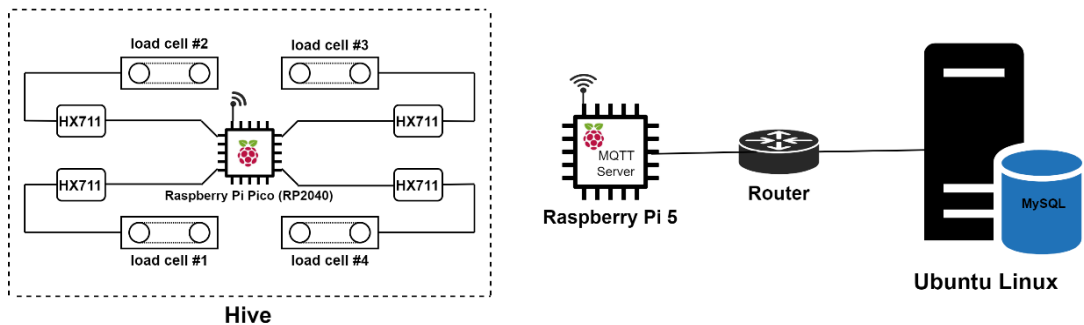

**Figure 4:** Laboratory scheme of the monitoring system.

Load cells that measure mechanical loads are connected to a 24-bit HX711 analogue-to-digital converter. The four HX711s are connected to an RP2040 controller based on a Raspberry Pi Pico board, which acts as a data processor at the primary stage.

The Raspberry Pi Pico collects and transmits digitized data from the load cells via a wireless Wi-Fi network using the MQTT protocol to a Raspberry Pi 5 microcomputer. The microcomputer acts as a central hub for collecting and processing information. At this stage, the obtained data is further processed according to formulas (5) and (6).

The processed data from the Raspberry Pi 5 is written to a MySQL database, which is hosted on a server running Ubuntu Linux. MySQL is used to store, organize and access data in real-time.

This system ensures continuous monitoring and data collection with sufficient accuracy.

As a result of the experiment on modelling bee colony movement, a database of measurements of bee colony movement in the hive during the winter season was formed.

Figure 5 shows a visualization of the movement of the bee colony according to the collected data.

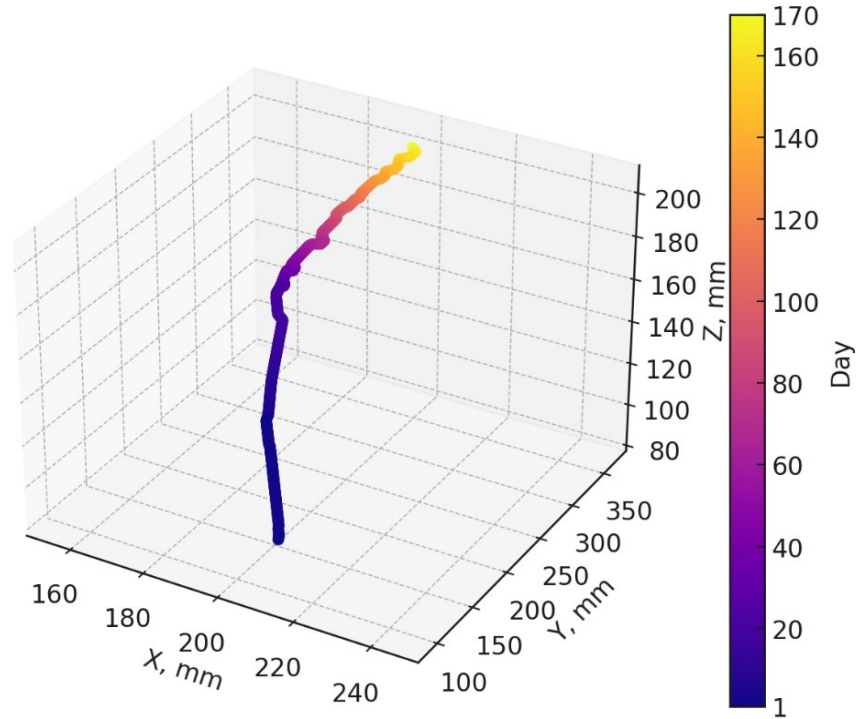

**Figure 5:** The graph of movement of the center of the bee family in the hive during the winter season according to the experimental modelling data.

The algorithm for finding the location of the bee colony in the hive works correctly, and the trajectory of movement based on the data from the load cells repeats the physical movement in the experiment, which simulates the movement of the bee colony in real conditions.

## 4. Conclusion

In this work, we have experimentally modelled the movement of a bee colony in a hive during the winter season and developed a monitoring system that allows tracking the movement of the bee colony and honey consumption without interfering with the natural activities of the bees. The proposed system is useful for beekeepers as it allows continuous monitoring of the bee colony during the winter period without the need to open the hive, which reduces the risk of stress for the bees.

Further research will be aimed at improving the system by adding additional sensors to measure other parameters of bee life, such as temperature, humidity and sound signals. This will allow us to get a more complete picture of the state of the bee colony and respond to possible problems in a timely manner.

# References

[1] Pita-Calvo C, Vázquez M. Honeydew Honeys: A Review on the Characterization and Authentication of Botanical and Geographical Origins. J Agric Food Chem (2018) 2523-2537. doi: 10.1021/acs.jafc.7b05807.

[2] Ng WJ, Sit NW, Ooi PA, Ee KY, Lim TM. The Antibacterial Potential of Honeydew Honey Produced by Stingless Bee (Heterotrigona itama) against Antibiotic Resistant Bacteria. Antibiotics (Basel) (2020) 871. doi: 10.3390/antibiotics9120871.

[3] Seeburger VC, D'Alvise P, Shaaban B, Schweikert K, Lohaus G, Schroeder A, Hasselmann M. The trisaccharide melezitose impacts honey bees and their intestinal microbiota. PLoS One (2020). doi: 10.1371/journal.pone.0230871.

[4] Seeley, Thomas & Camazine, Scott & Sneyd, James. Collective decision-making in honey bees: How colonies choose among nectar sources. Behavioral Ecology and Sociobiology (1991) 277-290. doi: 10.1007/BF00175101.

[5] Douglas S. Kridi, Carlos Giovanni N. de Carvalho, Danielo G. Gomes, Application of wireless sensor networks for beehive monitoring and in-hive thermal patterns detection, Computers and Electronics in Agriculture (2016) 221-235. https://doi.org/10.1016/j.compag.2016.05.013.

[6] Olivier Debauche, Meryem El Moulat, Saïd Mahmoudi, Slimane Boukraa, Pierre Manneback, Frédéric Lebeau, Web Monitoring of Bee Health for Researchers and Beekeepers Based on the Internet of Things, Procedia Computer Science (2018) 991-998. https://doi.org/10.1016/j.procs.2018.04.103.

[7] Armands Kviesis, Aleksejs Zacepins, System Architectures for Real-time Bee Colony Temperature Monitoring, Procedia Computer Science (2015) 86-94. https://doi.org/10.1016/j.procs.2014.12.012.

[8] Víctor Sánchez, Sergio Gil, José M. Flores, Francisco J. Quiles, Manuel A. Ortiz, Juan J. Luna, Implementation of an electronic system to monitor the thermoregulatory capacity of honeybee colonies in hives with open-screened bottom boards, Computers and Electronics in Agriculture (2015) 209-216. https://doi.org/10.1016/j.compag.2015.10.018.

[9] Nicolás Pérez, Florencia Jesús, Cecilia Pérez, Silvina Niell, Alejandro Draper, Nicolás Obrusnik, Pablo Zinemanas, Yamandú Mendoza Spina, Leonidas Carrasco Letelier, Pablo Monzón, Continuous monitoring of beehives' sound for environmental pollution control, Ecological Engineering (2016) 326-330. https://doi.org/10.1016/j.ecoleng.2016.01.082.

[10] D.R. Reynolds, J.R. Riley, Remote-sensing, telemetric and computer-based technologies for investigating insect movement: a survey of existing and potential techniques, Computers and Electronics in Agriculture (2002) 271-307. https://doi.org/10.1016/S0168-1699(02)00023-6.

[11] Chiu Chen, En-Cheng Yang, Joe-Air Jiang, Ta-Te Lin, An imaging system for monitoring the in-and-out activity of honey bees, Computers and Electronics in Agriculture (2012) 100-109. https://doi.org/10.1016/j.compag.2012.08.006.

[12] Joe-Air Jiang, Chien-Hao Wang, Chi-Hui Chen, Min-Sheng Liao, Yu-Li Su, Wei-Sheng Chen, Chien-Peng Huang, En-Cheng Yang, Cheng-Long Chuang, A WSN-based automatic

monitoring system for the foraging behavior of honey bees and environmental factors of beehives, Computers and Electronics in Agriculture (2016) 304-318. https://doi.org/10.1016/j.compag.2016.03.003.
[13] Thi Nha Ngo, Kung-Chin Wu, En-Cheng Yang, Ta-Te Lin, A real-time imaging system for multiple honey bee tracking and activity monitoring, Computers and Electronics in Agriculture (2019). https://doi.org/10.1016/j.compag.2019.05.050.
[14] A.E. Souza Cunha, J. Rose, J. Prior, H.M. Aumann, N.W. Emanetoglu, F.A. Drummond, A novel non-invasive radar to monitor honey bee colony health, Computers and Electronics in Agriculture (2020). https://doi.org/10.1016/j.compag.2020.105241.
[15] Thu Huong Truong, Huu Du Nguyen, Thi Quynh Anh Mai, Hoang Long Nguyen, Tran Nhat Minh Dang, Thi-Thu-Hong Phan, A deep learning-based approach for bee sound identification, Ecological Informatics (2023). https://doi.org/10.1016/j.ecoinf.2023.102274.
[16] Fiona Edwards-Murphy, Michele Magno, Pádraig M. Whelan, John O'Halloran, Emanuel M. Popovici, b+WSN: Smart beehive with preliminary decision tree analysis for agriculture and honey bee health monitoring, Computers and Electronics in Agriculture (2016) 211-219. https://doi.org/10.1016/j.compag.2016.04.008.
[17] Rahman Tashakkori, Abdelbaset S. Hamza, Michael B. Crawford, Beemon: An IoT-based beehive monitoring system, Computers and Electronics in Agriculture (2021). https://doi.org/10.1016/j.compag.2021.106427.
[18] Abdelbaset S. Hamza, Rahman Tashakkori, Bejamen Underwood, William O'Brien, Chris Campell, BeeLive: The IoT platform of Beemon monitoring and alerting system for beehives, Smart Agricultural Technology (2023). https://doi.org/10.1016/j.atech.2023.100331.
[19] Sahin Aydin, Mehmet Nafiz Aydin, Design and implementation of a smart beehive and its monitoring system using microservices in the context of IoT and open data, Computers and Electronics in Agriculture (2022). https://doi.org/10.1016/j.compag.2022.106897.
[20] Duarte Cota, José Martins, Henrique Mamede, Frederico Branco, BHiveSense: An integrated information system architecture for sustainable remote monitoring and management of apiaries based on IoT and microservices, Journal of Open Innovation: Technology, Market, and Complexity (2023). https://doi.org/10.1016/j.joitmc.2023.100110.
[21] Schmidt, Albrecht & Strohbach, Martin & Van Laerhoven, Kristof & Gellersen, Hans. Ubiquitous Interaction — Using Surfaces in Everyday Environments as Pointing Devices. Lecture Notes in Artificial Intelligence (Subseries of Lecture Notes in Computer Science) (2002) 263-279. doi: 10.1007/3-540-36572-9_21.
[22] K. Murao, J. Imai, T. Terada, & M. Tsukamoto, Activity Recognition and User Identification based on Tabletop Activities with Load Cells. Journal of Information Processing (2017) 59–66. doi:10.2197/ipsjjip.25.59.
[23] P.K. Shtanko, V.I. Shevchenko, O.S. Omelchenko, L.F. Dziuba, V.R. Pasika, O.M. Polyakov, Theoretical Mechanics: a textbook, National University of Zaporizhzhia Polytechnic, Zaporizhzhia, 2021.

[24] William Hesbach: Winter Management. Bee Culture Magazine, 2016. URL https://www.beeculture.com/winter-management.

[25] Main Department of the State Service of Ukraine for Food Safety and Consumer Protection in Kherson region, How to organize wintering of bees, 2019. URL https://dpss-ks.gov.ua/novini/yak-pravilno-organizuvati-zimivlyu-bdzhil.

[26] Main Directorate of the State Service of Ukraine for Food Safety and Consumer Protection in Kherson Oblast, Nutrition of bee colonies in winter, 2019. URL https://dpss-ks.gov.ua/novini/xarchuvannya-bdzholinix-simej-vzimku